# Digital image processing to detect subtle motion in stony coral

Shuaifeng Li[1], Liza M. Roger[2], Lokander Kumar[3], Nastassja Lewinski[2], Judith Klein[4], Alex Gagnon[5], Hollie M. Putnam[6], Jinkyu Yang[1]

[1]Department of Aeronautics and Astronautics, University of Washington, Seattle, Washington, 98195-2400, USA

[2]Chemical and Life Science and Engineering, Virginia Commonwealth University, Richmond, VA, United States

[3]Department of Physics, Colorado School of Mines, CO, USA

[4]Department of Chemistry, Colorado School of Mines, CO, USA

[5]School of Oceanography, University of Washington, Seattle, WA, USA

[6]Department of Biological Sciences, University of Rhode Island, Kingston, RI, 02881, USA

**Abstract**

Coral reef ecosystems support significant biological activities and harbor huge diversity, but they are facing a severe crisis driven by anthropogenic activities and climate change. An important behavioral trait of the coral holobiont is coral motion, which may play an essential role in feeding, competition, reproduction, and thus survival and fitness. Therefore, characterizing coral behavior through motion analysis will aid our understanding of basic biological and physical coral functions. However, tissue motion in the stony scleractinian corals that contribute most to coral reef construction are subtle and may be imperceptible to both the human eye and commonly used imaging techniques. Here we propose and apply a systematic approach to quantify and visualize subtle coral motion across a series of light and dark cycles in the scleractinian coral *Montipora capricornis*. We use digital image correlation and optical flow techniques to quantify and characterize minute coral motions under different light conditions. In addition, as a visualization tool, motion magnification algorithm magnifies coral motions in different frequencies, which explicitly displays the distinctive dynamic modes of coral movement. We quantified and compared the displacement, strain, optical flow, and mode shape of coral motion under different light conditions. Our approach provides an unprecedented insight into micro-scale coral movement and behavior through macro-scale digital imaging, thus offering a useful empirical toolset for the coral research community.

**Introduction**

Reef-building corals, as keystone organisms, support diverse marine communities and provide a host of ecosystem functions. They are composed of coral organisms living in symbiosis with photosynthetic dinoflagellate algae and a complex of additional bacterial, archaeal and fungal communities[1]. Coral behavior, physiology and ecology are impacted by anthropogenic global climate change through the global rise in sea-surface temperatures and ocean acidification[2]. Simultaneously, coral reefs have experienced substantial decline due to disease outbreaks[3], overfishing[4], coastal development and associated runoff[5]. The increasing frequency of marine heat waves has also resulted in mass coral mortality[6,7]. The combination of these local and global stressors is threatening coral reefs at an unprecedented scale.

One notable aspect of corals is their dynamic motion, which may be affected by the light condition, temperature, pH and other environmental variables[8–15]. Some soft corals, such as the family of *Xeniidae,* exhibit a unique, rhythmic pulsation, which is believed to enhance photosynthesis[16,17]. However, compared with the soft corals that usually have flexible extended tentacles, the scleractinian corals responsible for the framework structure of most reefs mineralize a rock-like skeleton made of calcium carbonate. In many species of scleractinian corals, the motion of tissue is more subtle, or imperceptible. Therefore, visualizing and quantifying the dynamic motion in order to extract information on coral behavior remains challenging for researchers studying Scleractinia. The dynamic motion of scleractinian corals can be meaningful and is worth being explored. For example, branched corals with small polyps will expand their tentacles to expose the photosynthetic symbionts to light during daytime. However, under the strong light, the coral will retract the tentacles to protect photosynthetic symbionts from irradiance[15,18,19]. Furthermore, some tentacles also serve as the probes to detect and kill competitors that settle within the wide aggressive reach of these massive corals[20]. The ciliary motion of tentacles also contributes to the mass transfer[21–23]. In areas away from coral polyps and their tentacles, the coenosarc, more subtle tissue movements have been revealed. These waves of tissue movement may speculatively be involved in enveloping or pumping seawater to different reservoirs within a coral[24]. Recently, seawater exchange rates in a growing coral were found to respond to stressful conditions like extreme ocean acidification[25]. Taken together, experiments like these hint at the rich connections that may exists between tissue movement, physiology, and how corals respond to environmental changes. New techniques that can image and quantify subtle tissue movements could uncover the role of tissue motion in coral health and help to better understand coral biophysical interactions and thus the fate of coral reefs in a changing ocean.

Imaging is a powerful way to provide information about the time-varying nature of the world. Photogrammetry microscopic imaging, and time-lapse imaging have resulted in the capacity of detecting change at both reef[26,27] and cellular levels[28–30]. New imaging techniques borrowed from other fields, such as mechanics, aerospace engineering and biological engineering, promise even more detailed and quantitative information that could be applied to address the fine-scale analysis of coral movement under a changing environment[31,32]. For example, correlation-based image registration and tracking methods such as digital image correlation (DIC) and particle image velocimetry (PIV) measure the mechanics of materials and fluids[31,33], and thus could be used to map movement of the coral tissue surface[22]. Optical flow is another effective method to demonstrate the movement between the camera and moving objects[32]. These approaches can potentially quantify the pixel-level motion in terms of displacement and velocity in biological fields. To provide the evident motions, previous attempts on magnifying subtle and imperceptible movements have been made along two perspectives: *Lagrangian* perspective and *Eulerian* perspective[34–37]. For instance, the imperceptible change in face color can be magnified to estimate the heart rate[38] and the functional role of the tectorial membrane in mammalian hearing can be revealed[39]. These approaches have rarely been used in marine biological systems including corals where they could quantify and visualize motion and further build a bridge between imaging techniques and biomechanics.

Here, we selected one species of hard coral, *Montipora capricornis*, to study the dynamic motion of coral polyps and coenosarc under diel cycling light conditions using time lapse imaging. We first quantify the coral motion by DIC and extract barely visible mechanical quantities in terms of displacements and strains. Next we present the optical flow to show the velocity field and motion polarization. These two methods offer physiological information in corals with high consistency, which could not have been sensed by naked eyes. Typical modes of motion are visualized by the phase-based motion magnification, which clearly shows a pattern of motion related to the corals' sensitivity to the changing light conditions. We provide a systematic and quantitative approach for characterizing and analyzing subtle and/or imperceptible coral tissue movements, thereby opening a new vista in coral behavioral, physiological, and cellular analysis.

Results

**Experimental observation of *Montipora capricornis*.** An aquarium experimental setup is used to observe the coral *Montipora capricornis* (Fig. 1a). It includes a digital single lens reflex (DSRL) camera with a macro lens to take pictures of the fine structures of coral tissue surface. To study the effect of light on coral tissue motion, aquarium light and ceiling light were used to mimic light conditions during the day and the night. 200 pictures are taken for each light condition with a rate of 30 frames per hour. Experimental setup details are shown in Supplementary figure 1 and are described in the Methods section.

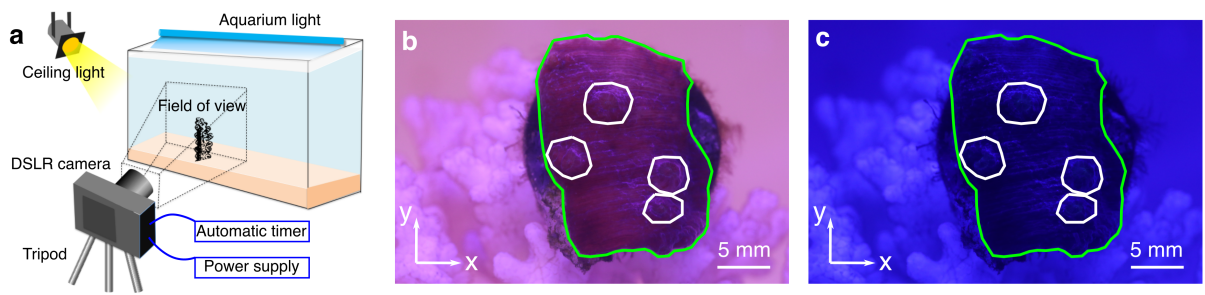

**Fig. 1 | Experimental observation of *Montipora capricornis*. a.** Schematic of experimental setup of a DSLR camera, aimed at the corals in the aquarium. **b.** Daylight pictures were taken under white light, while **c.** nighttime pictures were taken under blue light. The green curves in **b** and **c** enclose the coral tissue surface. The white curves in **b** and **c** enclose the polyps. Scale bars are shown in **b** and **c**.

Fig. 1b and Fig. 1c display the pictures taken at day and at night, respectively. The nearly planar arrangement of the coral provided the opportunity to obtain motion information from all polyps on the coral's surface. The heterogeneity of the tissue on surface creates texture that is sufficient for applying DIC and optical flow techniques. Luminance (as calculated and shown in Supplementary note 1 and Supplementary figure 2) was smaller in the blue light at night than at day, as the hue is more blueish at night due to the lack of the ceiling light. Visualization of coral dynamic motion, not perceptible by the human eye, can be observed from the time-lapse video played in 10 frames per second (×1200 speed; Supplementary movie 1).

**Characterization by digital image correlation.** In order to quantify coral tissue motion from a biomechanics perspective, we used a DIC technique to characterize the deformation of the surface relative to an initial reference picture (the first picture). Displacements ($u_x$) along horizontal direction between the reference picture and the second picture at day and night (Fig. 2a). The strongest motions are concentrated around polyps and the edge of the fragment. At night, most of the coral tissues surface undergoes horizontal motion, while few coral tissues are moving horizontally at day. Displacements ($u_y$) along the vertical direction were also quantified between the reference picture and the second picture at day and night (Fig. 2b). Similar to the distribution of displacement $u_x$, the strongest motions are concentrated around polyps and the edge of the fragment. Vertical tissue motion is more widespread across the coral at night compared to daytime. The displacement of both $u_x$ and $u_y$, in comparison to the reference picture between day and night, can be seen in videos created from the 200 pictures (Supplementary movie 2 and Supplementary movie 3). These visualizations highlight that the coral tissue surface is moving dynamically over time, with more movement observed during the night than during the day.

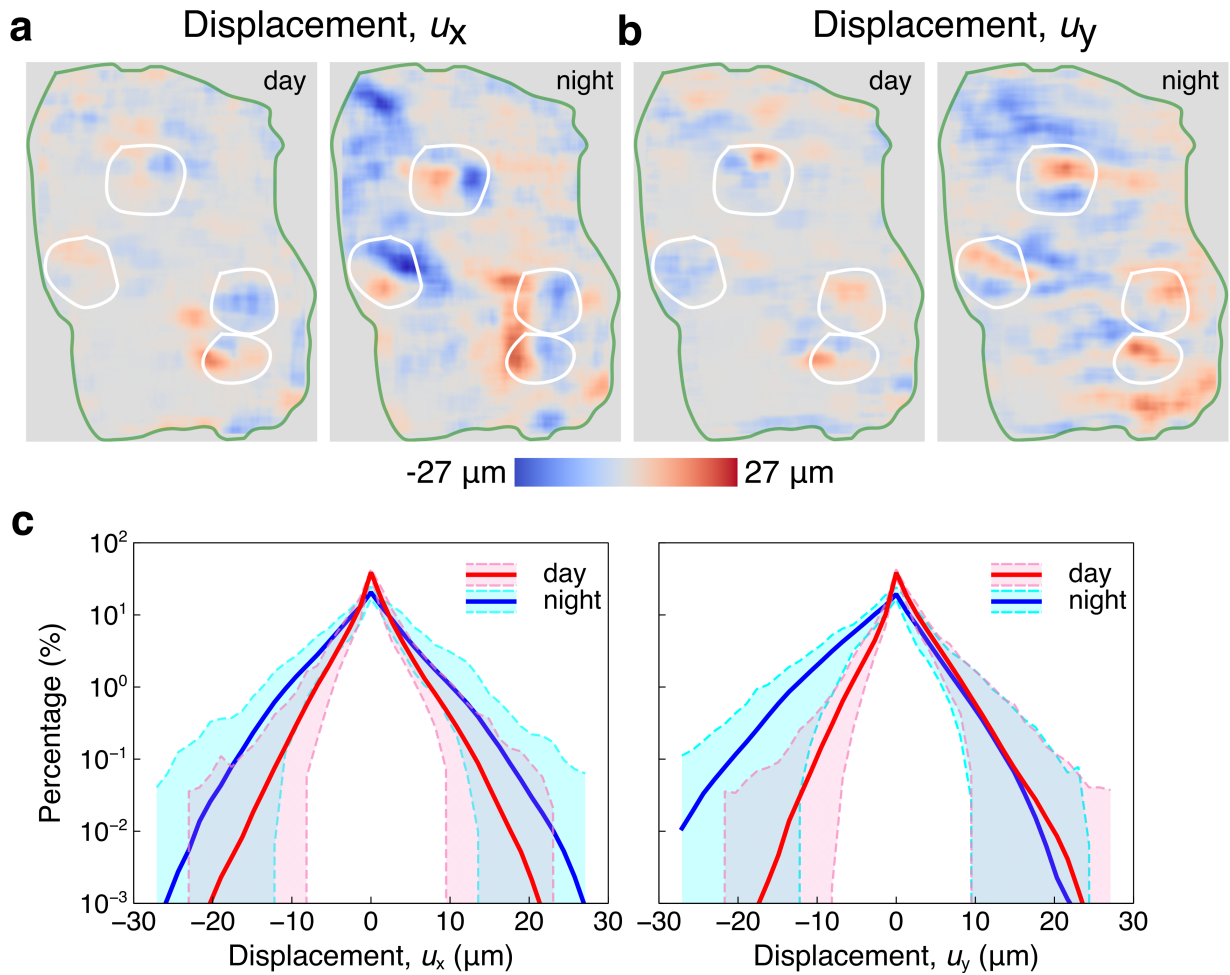

**Fig. 2 | Displacements of the coral tissue surface from digital image correlation. a.** The displacements $u_x$ along horizontal direction between the reference picture and the second picture at day and night are shown in the left and right panel, respectively. **b.** The displacement $u_y$ along vertical direction between the reference picture and the second picture at day and night are shown in the left and right panel, respectively. Green curves in **a** and **b** represents the edge of the fragment and white curves enclose the polyps. **c.** The percentage histograms of displacement $u_x$ and displacement $u_y$ are shown in the left and right panels, respectively. The red and blue curves indicate the means of each displacement during day and night. The pink and cyan regions are variations during day and night, respectively.

Across ~13 hours, histograms of displacements $u_x$ and $u_y$ are continuously changing (Supplementary movie 4). Most of the areas of the coral tissue surface are barely moving, i.e. nearly zero displacement (Fig 2c). For the portions in motion, more areas of the coral tissue surface are contributing to the motion at night since the blue curves are above the red curves, coinciding with the displacements shown in Fig. 2a and Fig. 2b. Furthermore, the histogram of displacement $u_x$ (Fig. 2c) illustrates the nearly symmetric distribution, suggesting that there is no motion preference and the system is balanced along the horizontal direction. However, the histogram of displacement $u_y$ shows an asymmetric distribution. During the day, the coral tissue surface tends to move upwards because there are more positive values, while at night, the tissue tends to move downwards, which indicates that motion preference based on the light condition is shown along the vertical direction.

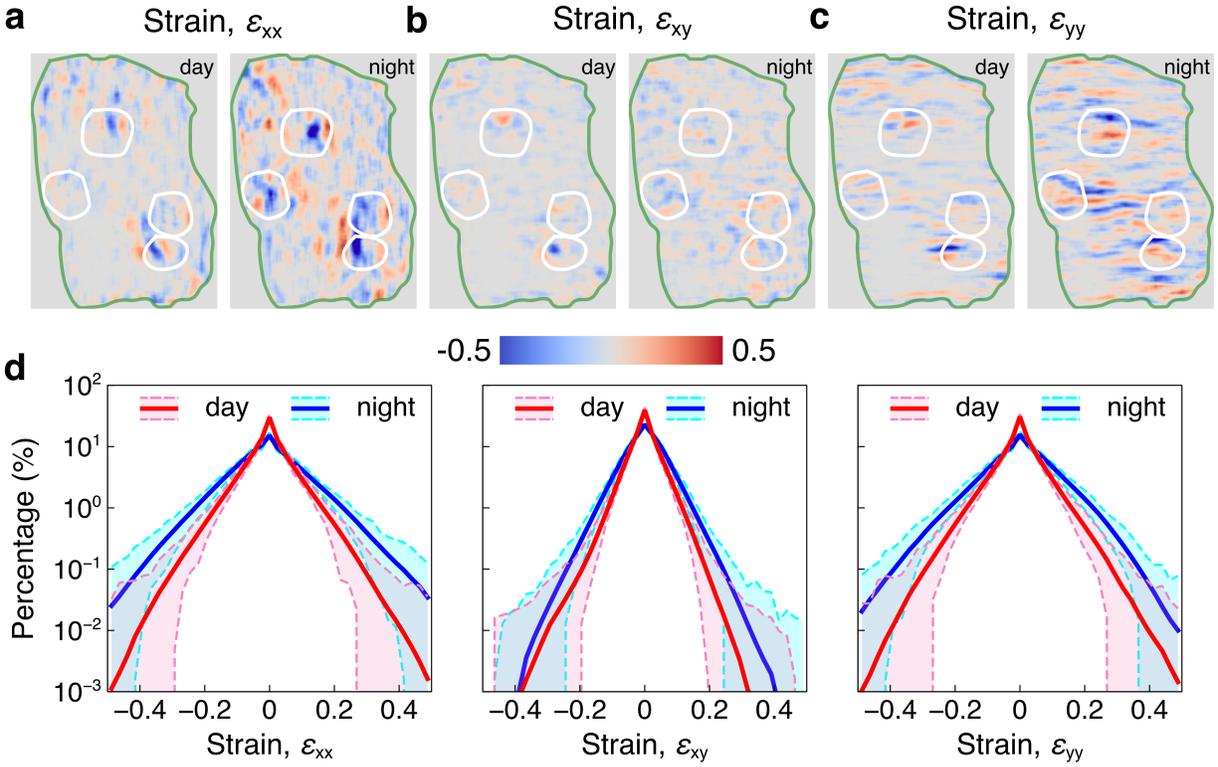

**Fig. 3 | Strains of the coral tissue surface from digital image correlation. a-c.** Strains $\varepsilon_{xx}$, $\varepsilon_{xy}$ and $\varepsilon_{yy}$ between the first and the second picture from day and night are shown in the left and right panel, respectively. Green curves in **a** and **b** represents the edge of the fragment and white curves enclose the polyps. **d.** The percentage histogram of the strains $\varepsilon_{xx}$, $\varepsilon_{xy}$ and $\varepsilon_{yy}$ are shown in the left, middle and right panels, respectively. The red and blue curves indicate the means of each strain during day and night. The pink and cyan regions are variations during day and night.

Strain measures the deformation of the material to identify whether it is under tension or compression. The linear strain is considered here. To obtain the accurate strain, smoothing and noise reduction of displacement field are necessary. We adopt the point-wise local least-square fitting technique[40,41]. The strain calculation window is set to 19×19. Supplementary movie 5 shows the normal strain $\varepsilon_{xx}$ and $\varepsilon_{yy}$, shear strain $\varepsilon_{xy}$ during day and night, respectively. The largest strain is concentrated around the polyps and obviously the distribution of $\varepsilon_{xx}$ is more evident at night (Fig. 3a), supporting that more of the coral tissue surface participate in the motion at that time. Shear strain ($\varepsilon_{xy}$) and normal strain ($\varepsilon_{yy}$) (Fig. 3b and Fig. 3c) show similar patterns. Compared with the $\varepsilon_{xx}$, $\varepsilon_{yy}$ has clearer alignment along the wrinkles on the coral surface, indicating the anisotropy of the coral tissue surface motion. Most of the coral tissue was not either under tension or compression due to the nearly zero strains, which is corresponding to the nearly zero displacements in the above analysis. Corresponding to the displacement, strain measurements also show larger values at night.

We note that we have also conducted a noise effect study to test the statistical significance of the DIC results presented above. Although correlation functions act to normalize images during image processing, quantification of noise effects provide a further verification of the observed patterns[42]. The enclosed coral skeleton on which our coral is fixed was analyzed by DIC (Supplementary Figure 3) for displacement (Supplementary Figure 4) and strain (Supplementary Figure 5). Details and analysis are in the Supplementary note 2. The results indicate that the DIC results are valid, since the noise is smaller than the extracted signal (Supplementary Figure 4 and Supplementary Figure 5; see also Supplementary note 2 and Supplementary movie 6).

**Characterization by optical flow.** Measuring optical flow is an effective method to explore the pixel-wise motion information such as velocity. In order to increase the signal-to-noise ratio, we assume the motion field is constant in a small window (21×21) around each pixel and the smoothing kernel (19×19) is used for reducing noise. The basic optical flow equation is solved for all pixels in the window by the least squares criterion, which is also known as the Lucas-Kanade method[43]. Similar to DIC, for each case, we take the first picture as the reference and calculate the optical flow for the 200 pictures (Fig. 4). Supplementary movie 7 shows the velocities of the optical flow during day and night. The high velocity areas (high color saturation) are around the polyps (marked by black crosses) and around the tissue margins, whereas colors with low saturation are distributed around the coral tissue surface margin. At night, there are more colors with high saturation in the optical flow, again supporting our findings that coral movement is higher in magnitude at night.

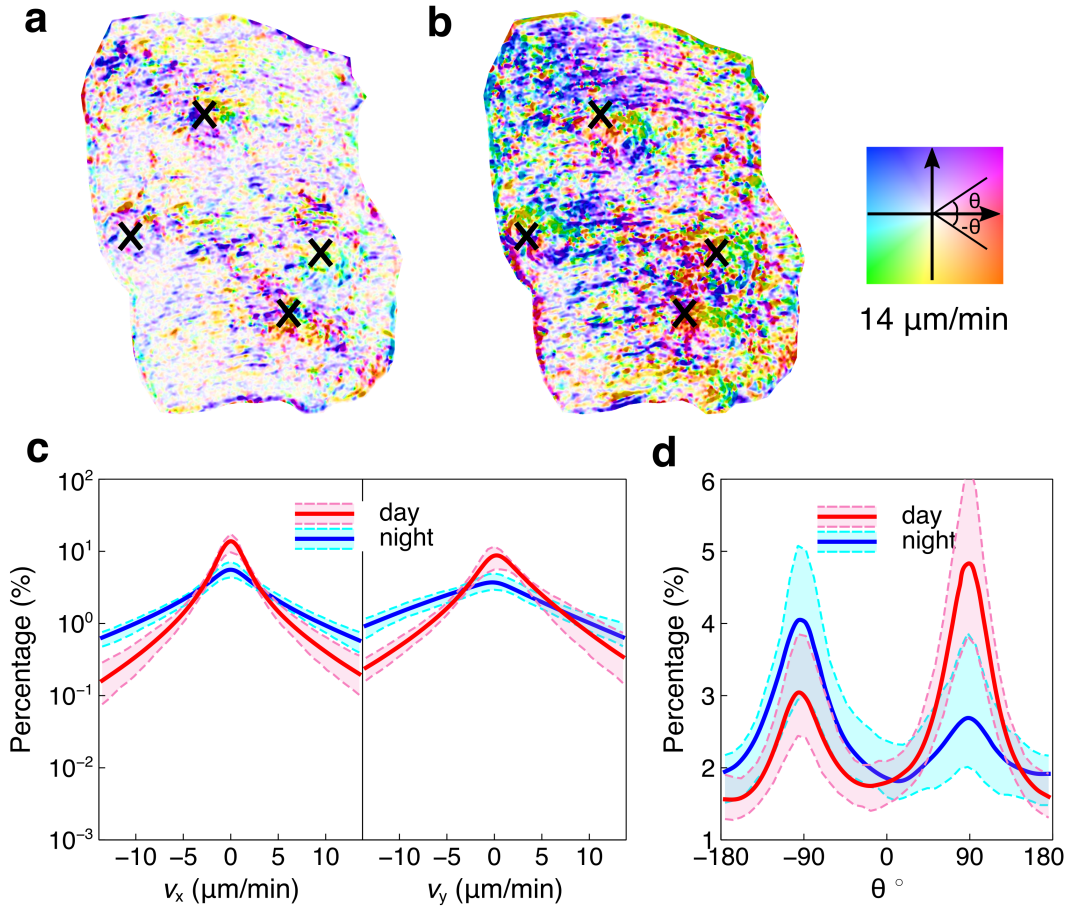

**Fig. 4 | Optical flow of the coral tissue surface. a-b.** Optical flow between the second and first pictures from day and night, respectively. Black crosses mark the position of polyps. **c.** The

percentage histograms of velocities $v_x$ and $v_y$ along horizontal and vertical directions. The red and blue curves indicate the means of each velocity from day and night. The pink and cyan regions are variations between day and night. **d.** The percentage histogram of direction of velocity. The red and blue curves indicate the means of each angle from day and night. The pink and cyan regions are variations during day and night.

Similar to the DIC results, along the horizontal direction, the histogram is symmetric (Fig. 4c), indicating that the coral tissue surface does not have statistically significant motion preference along the horizontal direction. The histogram of $v_y$ shows the similar trend with the displacement $u_y$ from DIC (Fig. 4c). During the day, the coral tissue surface tends to present a positive velocity translating upwards motion, while the tissue tends to move downwards with a negative velocity at night. This observation is even more evident in Fig. 4d, which offers a histogram of direction of optical flow with 90° and –90° represent the upward and downward direction, respectively. According to these measurements, the dominant motion is along the vertical direction that the light condition the coral is under, since there are evident peaks in 90° and –90°. Again there is a contrast in the motion between differing light conditions. In the day, there is a higher peak in 90°, which means that the tissue tends to move upwards, while at night, the higher peak in –90° suggests the downwards motion. Similarly, we conducted the noise effect study as shown in Supplementary figure 6. The results show that the velocities on the coral skeleton is smaller, thus support the validation of the analysis mentioned above. Details are described in Supplementary note 3.

**Motion magnification for the coral tissue surface.** Exploring the biological and physical modes of coral tissue motion is crucial to understand the coral behaviors. Here we further process the optical flow result to obtain the deformation of coral tissue surface at different frequencies and magnify the modes of this movement to make them qualitative and visible to human eyes. To this end, we performed the Fourier transform on the optical flow results and extract the absolute value of velocity in different frequencies. Zero padding is used to increase the frequency resolution. Supplementary movie 8 shows mode shapes of coral tissue in the frequency range of 0 $min^{-1}$ to 0.25 $min^{-1}$ (Nyquist frequency). As the frequency increases, the intensity of the motion decreases (Fig. 5). In the frequency 0.05 $min^{-1}$ during the day the motion is concentrated around the polyps and margins (Fig. 5a). In stark contrast, the mode shape in the same frequency in the dark is not only around the polyps and margins, but also on the coral tissue surface. Under both light conditions, a greater proportion of coral tissue contributes to the motion with the increasing frequency. Overall, the mode of extremely slow coral tissue motion is observed, which coincides with the phenomenon observed in DIC and optical flow.

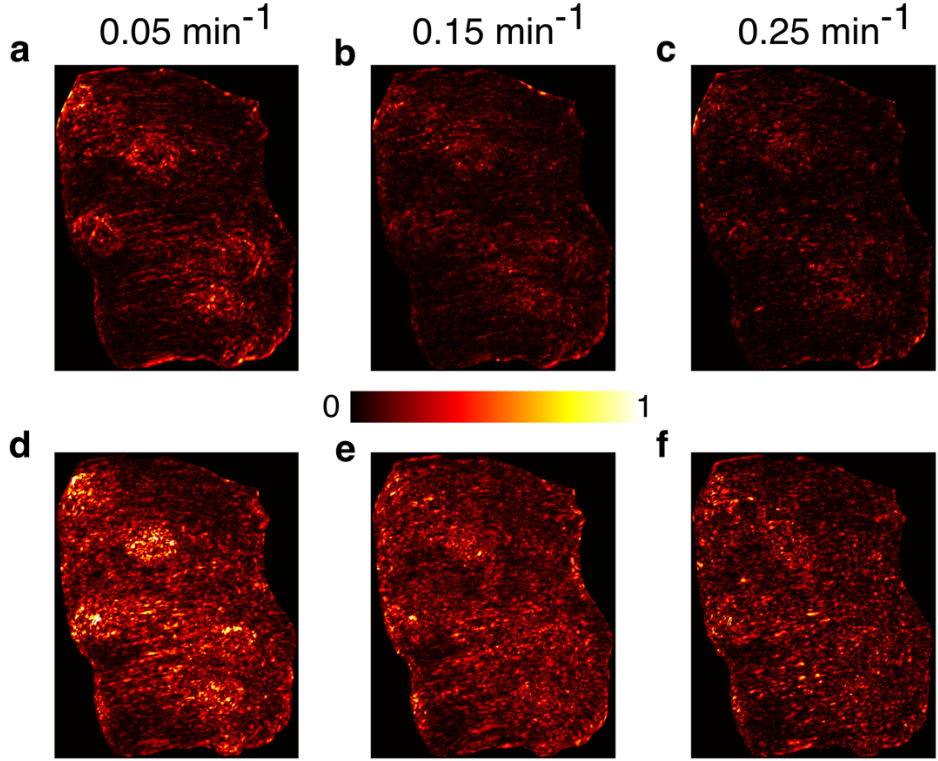

**Fig. 5 | Mode shape of the coral tissue motion during day and night. a-c.** Mode shapes of coral tissue motion during day at frequencies 0.05 min$^{-1}$, 0.15 min$^{-1}$ and 0.25 min$^{-1}$, respectively. **d-f.** Mode shapes of coral tissue motion during night at frequencies 0.05 min$^{-1}$, 0.15 min$^{-1}$ and 0.25 min$^{-1}$, respectively.

In order to visualize the mode shape in different frequencies, we used a phase-based motion magnification[16]. A type of image representation, the complex steerable pyramid, which is an over-complete transformation that can decompose an image according to spatial scale, orientation and position, is used to magnify the motion in a specific frequency range[44]. The image is completely decomposed into amplitudes $A_{\omega,\theta}(x,y)$ and phases $\varphi_{\omega,\theta}(x,y)$ at every scale $\omega$ and orientation $\theta$. We augment the motion 75 times by magnifying phase in the frequency range of interest. We choose three frequency ranges, 0.04 min$^{-1}$ ~ 0.06 min$^{-1}$, 0.14

min$^{-1}$ ~ 0.16 min$^{-1}$ and 0.22 min$^{-1}$ ~ 0.24 min$^{-1}$ to demonstrate the motion microscope for coral tissue mode shapes. Supplementary movie 9 shows the magnified video played in 10 frames per second, which is made of pictures taken during the day. The movie presents the polyps motion in the low frequency range and the coral tissue surface motion in the high frequency range. The motion magnitude is decreasing as the frequency increases. According to this analysis, more coral tissue is involved in the motion at night at these frequencies (Supplementary movie 10), which coincides with the mode shape analysis from the optical flow.

**Conclusion**

We applied digital image correlation, optical flow, and motion microscopy techniques to a series of images of *Montipora capricornis*. Though this hard coral has the slow and subtle tissue motion, which is hard to capture not only with naked eyes, but also with optical apparatus, we have successfully extracted its responsiveness to light conditions based on ~13 hours of high-frequency images. The blending of powerful and effective tools from mechanics and computer science and coral biology opens avenues for studying coral physiology from macro-scale pictures. This systematic approach offers us possibilities to quantify and visualize the subtle coral tissue surface motion and important physical and biological mode in a time-efficient, yet accurate manner. In this proof of concept study, we only focus on the effect of light to motion for a single coral, but the proposed approach can be generalized to study the effect of a variety of environmental variables on coral tissue motion.

**Methods**

**Aquarium maintenance.** The *Montipora capricornis* was bought from the local store (Seattle Corals Aquariums) and is raised in a 11.36 L aquarium with specific density = 1.024 g/cm$^3$ and pH = 8.4. Artificial seawater is made from Instant Ocean Reef Crystals Reef Salt with the salinity 34.1 ppt. A one third water change was carried out every 3 days to maintain steady aquarium conditions. Regular tests on pH, $NH_4^+$, $NO_2^-$ and $NO_3^-$ were done to make sure the suitable water quality for coral growth. The continuous water flow within the tank was provided by the Hydor Koralia Nano Aquarium Circulation Pump with 908.5 L per hour flow rate. The temperature of the water was controlled by the 50-Watt FREESEA submersible heater and was maintained in 25.5°C. Light was provided on a 14.5:9.5 light: dark cycle using a 6-Watt NICREW ClassicLED aquarium light and the ceiling light in the laboratory. The aquarium light emits the blue light with lumens 380 lm, and it is turned on all day, while the ceiling light emits the white light and can be controlled to be on or off manually which is turned on at day and turned off at night.

**Pictures acquisition.** We use Canon EOS 5D Mark IV and EF 100mm f/2.8L Macro IS USM and timer to take pictures for the *Montipora capricornis* every 2 minutes automatically. The parameters for the camera are: F11, ISO4000, 1/10s. The camera is put on the tripod to make it stable and ready for the long-period shooting. The macro lens is perpendicular to the wall of the aquarium to avoid the blurred effect caused by the refraction.

**Data Availability**

Data supporting the findings of this study and Matlab codes for digital image correlation, optical flow, and motion microscopy are available for downloads in Open Science Framework. DOI: 10.17605/OSF.IO/49KMH.

**Acknowledgments**

This work is supported by funding from the National Science Foundation. SL and JY (HDR: DIRSE-IL: 1939249), HMP (HDR: DIRSE-IL: 1939795), JKS (HDR: DIRSE-IL: 1940169), AG (OCE:CAREER: 1552694), and NL and LMR (HDR: DIRSE-IL: 1939699).

**Author contributions**

SL and JY proposed the research; SL conducted the experiments; SL performed the numerical analysis; HP and JY provided guidance throughout the research; SL, HP, and JY prepared the manuscript. LMR, LK, NL, JK, and AG provided guidance during manuscript preparation and development. All authors participated in manuscript editing.

**Additional information**

The authors declare that they have no competing financial interests. Correspondence and requests for materials should be addressed to Hollie M. Putnam (email: hputnam@uri.edu) and Jinkyu Yang (email: jkyang@aa.washington.edu).